\documentclass{PoS}

\pdfoutput = 1

\usepackage{amsmath}
\usepackage{amssymb}
\usepackage{graphicx}
\usepackage{cite}
\usepackage{subfig}
\usepackage{setspace}

\newcommand{\Fermi}[0]{\textit{Fermi}\xspace}
\def\ie{{\it i.e.}}
\def\Fermi{\,{\it Fermi}}

\title{Complementarity between collider, direct detection, and indirect detection experiments}

\ShortTitle{Complementarity between collider, direct detection, and indirect detection experiments}

\author{\speaker{Matthew Cahill-Rowley}%
         \thanks{A footnote may follow.}\\
        SLAC, USA\\
        E-mail: \email{mrowley@slac.stanford.edu}}

\abstract{
We examine the capabilities of planned direct detection, indirect detection, and collider experiments in exploring the 19-parameter p(henomenological)MSSM, focusing on the complementarity between the different search techniques. In particular, we consider dark matter searches at the 7, 8 (and eventually 14) TeV LHC, \Fermi, CTA, IceCube/DeepCore, and LZ. We see that the search sensitivities depend strongly on the WIMP mass and annihilation mechanism, with the result that different search techniques explore orthogonal territory. We also show that advances in each technique are necessary to fully explore the space of Supersymmetric WIMPs.}\FullConference{Science with the New Generation of High Energy Gamma-ray experiments, 10th Workshop \\
                 04-06 June 2014\\
                 Lisbon - Portugal}

\begin{document}

\section{Introduction}

Determining the identity of dark matter (DM) is one of the most pressing issues before us today. One promising class of dark matter candidates is Weakly Interacting Massive Particles (WIMPs), which predict the observed relic abundance through the simple mechanism of thermal freeze-out. WIMPs naturally appear in many extensions of the Standard Model (SM) that resolve the gauge hierarchy, with the most notable example being supersymmetry (SUSY). Several important classes of experimental techniques have been proposed to detect non-gravitational signatures of WIMP DM. These techniques include direct detection of WIMPs scattering off of nuclei, indirect detection of WIMPs by observing excesses of high-energy SM particles resulting from WIMP annihilation, and direct production of WIMPs in high energy colliders. In this paper, we seek to understand how these different techniques complement each other within the framework of the phenomenological Minimal Supersymmetric Standard Model (pMSSM). We find that the three techniques place orthogonal constraints on the parameter space and that advances in all three techniques are necessary to cover the supersymmetric WIMP sector. This paper presents results from the study described in~\cite{Cahill-Rowley:2014boa}. In particular, detailed descriptions of the pMSSM and the constraints we apply can be found in that document and the references contained therein.

It is well-known that R-parity conserving supersymmetry predicts a stable dark matter candidate in the form of the lightest SUSY particle (LSP). Cosmological observations require the LSP to have no electric or color charge. Models with the lightest neutralino, $\chi_1^0$, as the LSP satisfy these requirements and will be the focus of this study. The DM phenomenology of these models is determined not only by the composition of the LSP (whether it is mostly comprised of the superpartners of the U(1) or SU(2) gauge bosons or the neutral Higgses), but also in general on the other SUSY particles, which can alter the annihilation and scattering rates and are important for the model's discovery potential at the LHC. Unfortunately, the simplest SUSY scenario, the MSSM, has ($\sim$100) parameters, making it far too large to explore in full generality. However, many of these parameters are restricted by the non-observation of large flavor violating effects. This allows us to simplify the parameter space by imposing the following experimentally-motivated assumptions: ($i$) no new phase appearing in the soft-breaking parameters, \ie, CP conservation, ($ii$) Minimal Flavor Violation at the electroweak scale such that the CKM matrix drives flavor mixing, ($iii$) degenerate first and second generation 
soft sfermion masses, and ($iv$) negligible Yukawa couplings and associated A-terms for the first two generations. These assumptions reduce the original space down to the 19-parameter pMSSM. We emphasize that no assumption about high-scale physics, such as the mechanism of SUSY breaking or unification of sparticle masses, has been applied to produce the pMSSM, and that it is therefore an ``unprejudiced'' approach to understanding TeV-scale supersymmetry.

Despite these simplifications, 19 parameters is too large for a systematic grid approach. We therefore perform a random sample of the pMSSM, testing 3 million points against experimental and theoretical constraints. The result is 223256 parameter space points (which we will call ``models'') satisfying all pre-LHC experimental constraints. Note that only about 20\% of the models predict the correct Higgs mass within the calculational uncertainty. However, we have found the LHC and DM constraints to be essentially dependent of $m_h$ for the range of Higgs masses in our model set. We assume that the LSP has its thermal relic abundance (calculated using micrOMEGAs 2.4~\cite{micromegas}, and discard models for which the predicted abundance is \textit{larger} than the upper limit from WMAP 7, $\Omega h^2 < 0.1234$, with the one-sided limit allowing for the possibility that other particles (such as the QCD axion) comprise the remainder of the DM. The remaining constraints are described in~\cite{Cahill-Rowley:2014boa}.  

\begin{figure}[htbp]
\centerline{\includegraphics[width=3.5in]{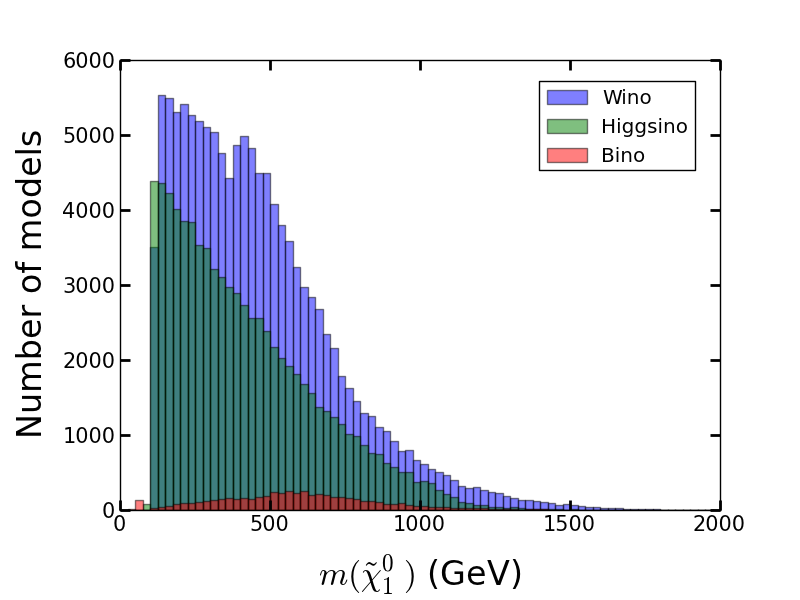}
\hspace{-0.50cm}
\includegraphics[width=3.5in]{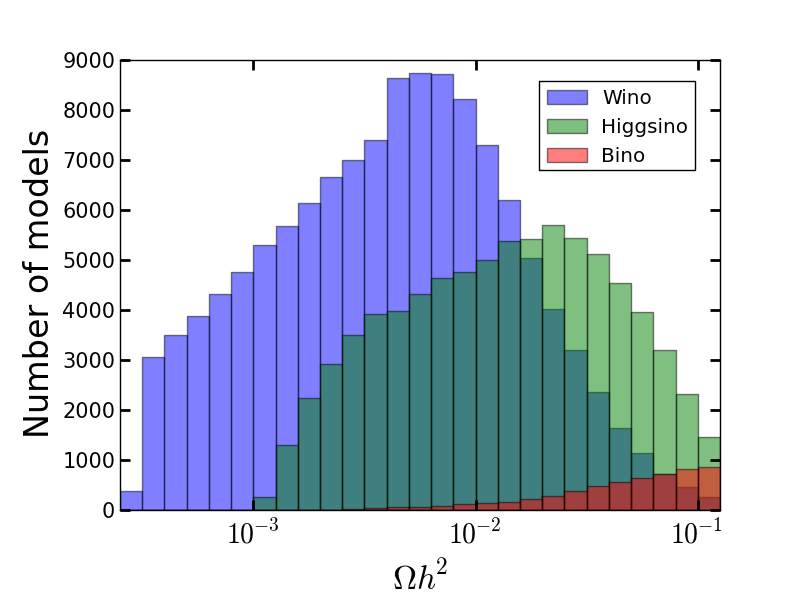}}
\vspace*{-0.10cm}
\caption{Distribution of the LSP masses (left) and predicted relic density (right) for the neutralino LSPs that are almost pure weak eigenstates in our model sample.}
\label{fig0}
\end{figure}

As a result of our scan ranges for the electroweak gauginos (chosen for compatibility with LEP data and to enable phenomenological studies at the 14 TeV LHC), the 
LSPs in our model sample are typically very close to being in a pure electroweak eigenstate as the off-diagonal elements of the chargino and neutralino mass matrices are 
at most $\sim M_W$. Figure~\ref{fig0} presents some properties of the nearly pure eigenstate LSPs (defined here as a single electroweak eigenstate comprising over 90\% of 
the mass eigenstate). The left panel displays the distribution of the LSP mass for nearly pure bino, wino, and Higgsino LSPs, while the right panel shows the 
corresponding distribution for the predicted LSP thermal relic density. Note that the LSP masses lie below $\sim 2$ TeV in all models; this is due to our choice of scan ranges as the entire SUSY spectrum must be lighter than $\sim 4$ TeV and heavier than the LSP (by definition), and this becomes increasingly improbable with increasing LSP mass. In addition, the relic density upper limit becomes increasingly difficult to satisfy at larger LSP masses. Similarly, due to LEP and relic density constraints, 
none of our models have LSP masses below $\sim$40 GeV. The fraction of models where the LSP is nearly a pure bino eigenstate 
is found to be rather low since such models lead to too high a value for the relic density unless they co-annihilate with another sparticle, happen to be close to a ($Z,h,A$) funnel region, or have a suitable Higgsino admixture. Note that only in the rightmost bin of the right panel is the relic density approximately saturating the WMAP/Planck thermal relic value.

\begin{figure}[htbp]
\centerline{\includegraphics[width=4.0in]{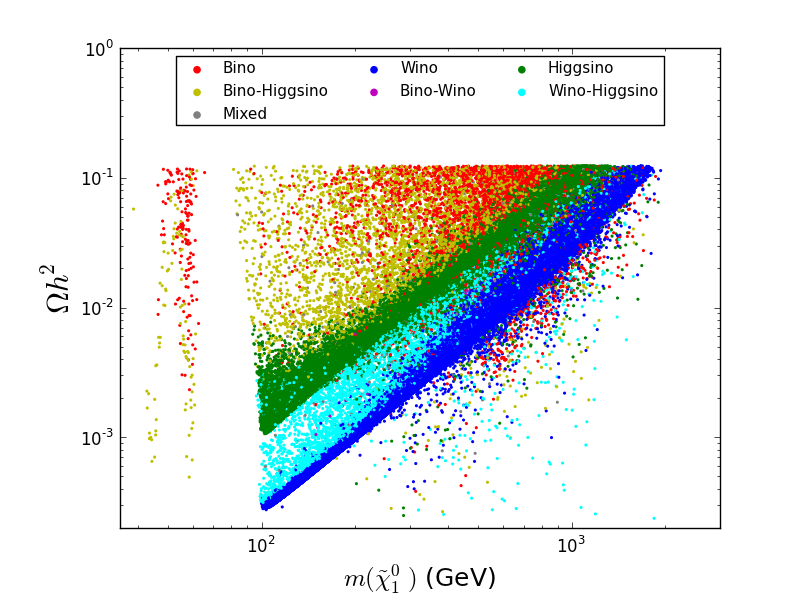}}
\vspace*{-0.10cm}
\caption{Thermal relic density generated in our pMSSM model set as a function of the LSP mass, color coded by the electroweak properties of the  
LSP as indicated and discussed in the text.}
\label{fig00}
\end{figure}

Figure~\ref{fig00} shows the thermal relic density as a function of the LSP mass, with model points color coded by their electroweak 
eigenstate content. We define ``pure'' LSPs as having a single eigenstate fraction $\geq 90\%$. Points shown as bino-wino, bino-Higgsino, or wino-Higgsino mixtures have 
less than $2\%$ Higgsino, wino, or bino fraction, respectively. ``Mixed'' points have no more than $90\%$ and no less than $2\%$ of each component. This plot clearly shows the different regions corresponding to different annihilation mechanisms: ($i$) The set of models with low LSP masses (forming `columns' on the left-hand side of the figure) correspond to bino-Higgsino admixtures which annihilate resonantly through the $Z,h$ funnels; note that these can be displayed as ``pure'' binos if the Higgsino fraction is below 10\%. ($ii$) The bino-Higgsino LSPs saturating the relic density in the upper-left region of the figure are of the so-called `well-tempered' variety. ($iii$) the pure bino models in the upper middle region of the Figure are bino co-annihilators (mostly with sleptons) or annihilate resonantly through the $A$ funnel. ($iv$) The green (blue) bands are pure Higgsino (wino) models that saturate the relic density bound (using perturbative calculations which do not include the Sommerfeld enhancement effect{\footnote {The Sommerfeld enhancement can significantly deplete the relic density of wino LSPs heavier than $\sim$ 1 TeV, while Higgsino and light wino LSPs are relatively unaffected~\cite{Hryczuk:2010zi}. Bino LSPs do not exhibit the effect because they can't exchange gauge bosons. Including the enhancement would increase the low-velocity annihilation cross section for heavy winos, lowering their predicted relic density but increasing their present-day annihilation cross section. Since the average velocity today is lower than during freeze-out, we would naively expect that including the enhancement would strengthen the limits on heavy wino LSPs. We will see that CTA is already able to exclude models with heavy winos in our perturbative calculation; we therefore expect that including the enhancement would minimally affect our conclusions.}}) near $\sim 1(1.7)$ TeV and have very low relic densities for lighter LSP masses. Wino-Higgsino hybrids are seen to lie between these two cases as expected. ($v$) A smattering of models with additional (or possibly multiple) annihilation channels are loosely distributed in the lower right-hand corner of the Figure. As we will see, many of the searches for DM are particularly sensitive to one or more of these LSP categories.

\section{LHC Searches}

We begin with a short overview of the constraints from 7 and 8 TeV LHC data. In order to get a comprehensive picture of the LHC's impact, we simulate 37 SUSY searches at the 7 and 8 TeV LHC, representing every relevant ATLAS SUSY search publicly available as of the beginning of March 2013, the more recent 20 fb$^{-1}$ 2-6 jets + MET 
analysis, the search for MSSM Higgses through di-tau production, and several CMS analyses. We find that the combined LHC searches exclude 45.5\% of the pMSSM models.    

\begin{figure}[htbp]
\centerline{\includegraphics[width=3.5in]{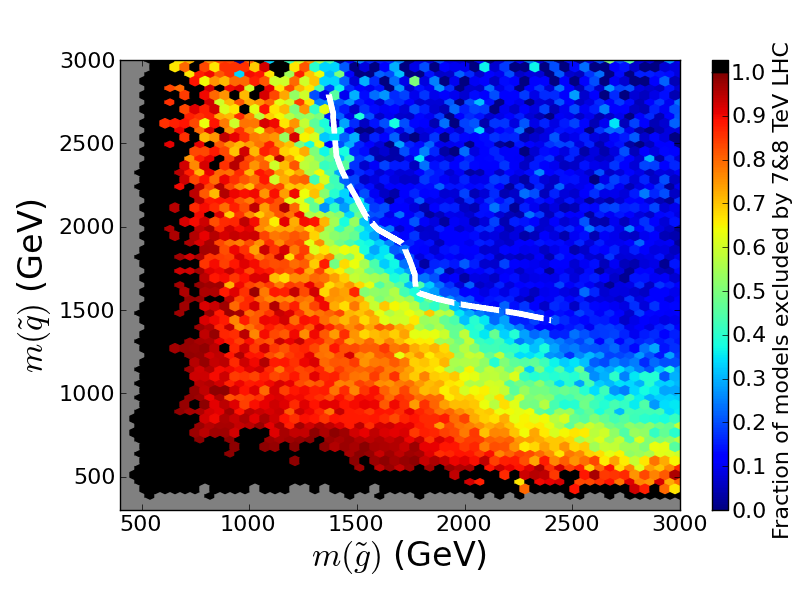}
\hspace{0.30cm}
\includegraphics[width=3.5in]{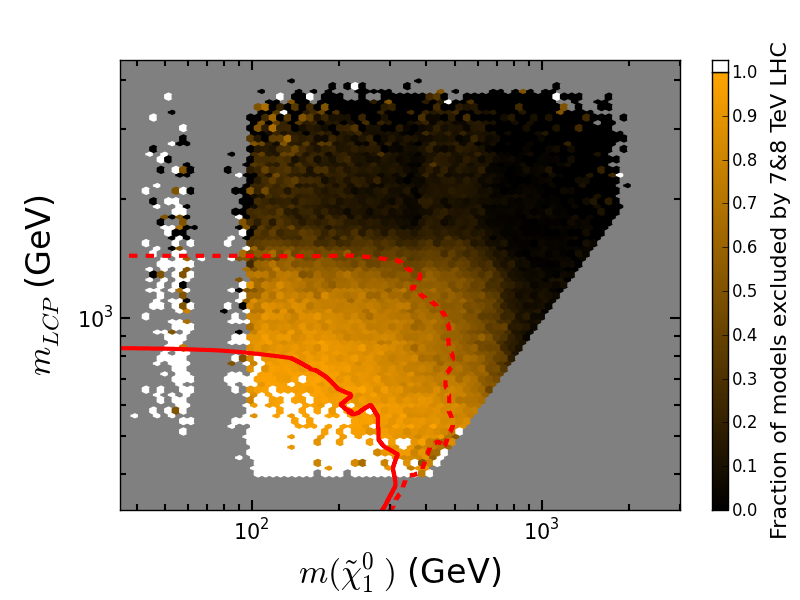}}
\vspace*{-0.10cm}
\caption{Top left: The fraction of pMSSM models excluded by the the 7 and 8 TeV LHC searches, shown in the lightest squark-gluino mass plane. The dashed line depicts the limit from the ATLAS 20 fb$^{-1}$ Jets+MET search. Top right: Fraction of models excluded in the LSP mass - lightest squark/gluino plane. Solid (dashed) lines show the simplified model limit from the ATLAS 20 fb$^{-1}$ Jets+MET search for degenerate squarks (the gluino) decaying directly to the LSP.}
\label{fig1}
\end{figure}

Currently, the sensitivity of the LHC comes mainly from its ability to produce colored sparticles with large rates. This is demonstrated by the left panel of Figure~\ref{fig1}, which shows the fraction of models excluded in the gluino-lightest squark mass plane. We see that the LHC excludes a large majority of models with light squarks or gluinos below $\sim$ 1 TeV, but only a small fraction of models for which both the squarks and gluinos are heavier than $\sim$ 1.8 TeV. The lesson here is that models with light sleptons, neutralinos, and charginos do not yet face strong constraints from LHC searches. The other key factor affecting the LHC's sensitivity is the LSP mass, as shown by the right panel of Figure~\ref{fig1}. We see that the fraction of models excluded drops precipitously as the LSP mass approaches the mass of the lightest colored sparticle, and that the fraction of models excluded with LSPs heavier than $\sim$ 700 GeV is very small. This is the well-known effect of spectrum compression - models with heavy LSPs produce soft decay products which are swamped by the large hadron collider backgrounds.

In addition to these results from the 7 and 8 TeV data, we also include projections for the results of 14 TeV null searches. We find that a combination of the 14 TeV jets+MET analysis and zero- and one-lepton stop analyses with 300 (3000) fb$^{-1}$ of data is expected to exclude 90.8\% (97.2\%) of models which have the correct Higgs mass and survive the 7/8 TeV searches.

\section{Direct Detection}

\begin{figure}[htbp]
\centerline{\includegraphics[width=3.5in]{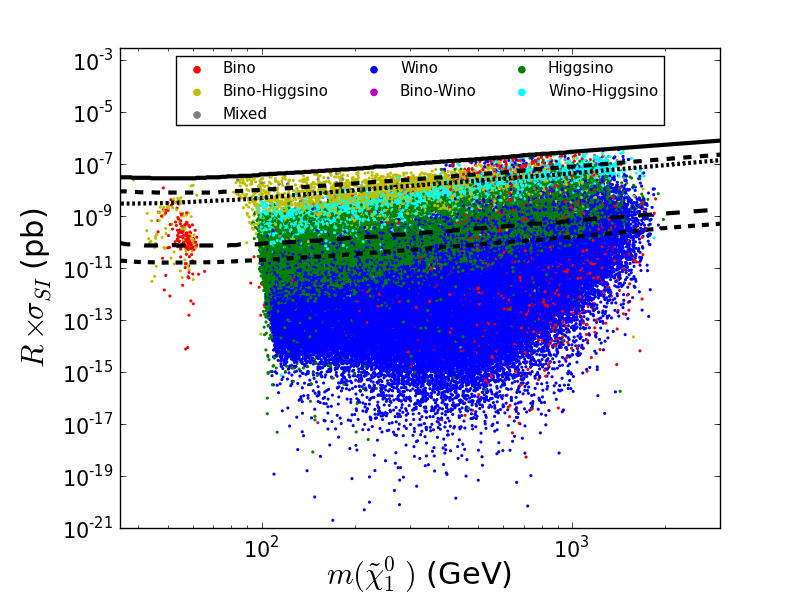}
\hspace{-0.50cm}
\includegraphics[width=3.5in]{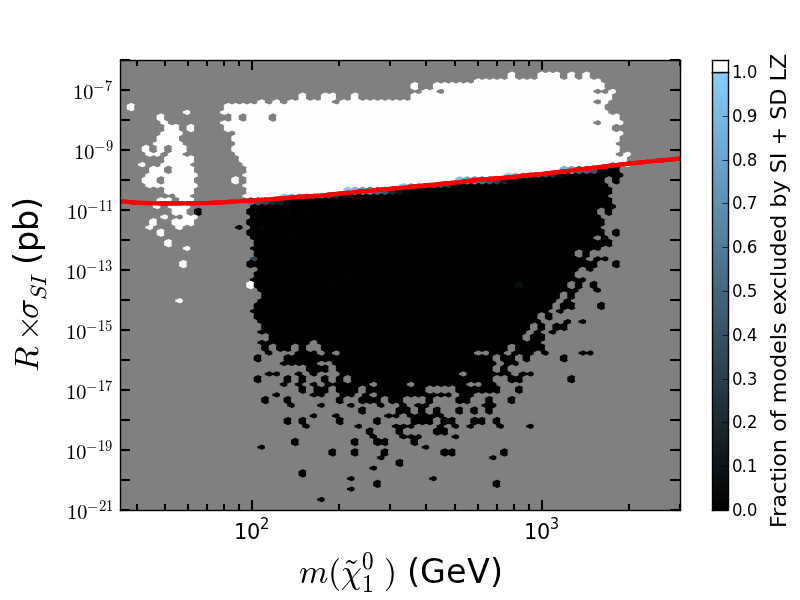}}
\vspace*{-0.10cm}
\caption{Scaled spin-independent cross sections for our pMSSM model set. The left panel shows the LSP composition, using the same color scheme as in Figure~2. The right panel is color coded by the fraction of models in each bin which could be excluded by SI+SD measurements with LZ. Black lines show the current or projected limits from Xenon 2011, Xenon 2012, LUX, Xenon 1T, and LZ, from top to bottom. The scaling factor R accounts for the possibility that the LSP abundance is less than the observed DM abundance.}
\label{fig3}
\end{figure}

WIMP dark matter is generally expected to have significant spin-independent (SI) or spin-dependent (SD) interactions with target nuclei, which can be detected by nuclear recoil experiments. We therefore compare the scattering cross sections predicted by micrOMEGAs with the expected limits from the LZ experiment. For this comparison, we rescale the scattering cross section by the LSP abundance (since the LSP is not necessarily all of DM), and weaken the expected limits by a factor of four to account for uncertainties in the scattering cross section from nuclear form factors. LSPs with sizable bino and Higgsino contents can have large couplings to the CP-even Higgs ($Z$) bosons, leading to large SI (SD) cross sections, respectively. LSPs that are mostly bino or wino can also scatter through squark exchange, which contributes to both SI and SD scattering. However, this scattering rate can be very small if the squark masses are large (as will be increasingly required by null LHC results). Models with very pure LSPs can therefore have very low direct detection cross sections. Figure~\ref{fig3} displays the SI cross section for our pMSSM models, color coded according to the LSP composition (left) and the fraction of models that would be excluded by null results from SI + SD scattering at the LZ experiment (right). From the left panel, we see that the well-tempered neutralinos are entirely within reach of LZ, while pure LSPs (particularly pure winos) can have an undetectably small cross section. In the right panel, we see that SD scattering is generally sensitive only to models that are expected to be excluded by SI scattering. The exception is the $Z/h$ funnel region, where SD scattering measurements are expected to exclude all of the models missed by SI scattering. Only a single model with a LSP lighter than $\sim$ 90 GeV is projected to survive both SI and SD searches; in this case the LSP is a highly pure bino that annihilates through a light stau.

\section{Indirect Detection}

Another important probe of our pMSSM models comes from indirect searches for the annihilation products of DM. We first consider the future impact of searches for gamma ray excesses in dwarf spheroidal galaxies by \Fermi~ and in the galactic center by CTA. We calculate the annihilation spectrum for each model using DarkSUSY 5.0.5~\cite{Gondolo:2004sc}, then compare it to the projected sensitivities. For \Fermi, we assume a 10-fold improvement in the sensitivity of the current dwarf analysis, resulting from increased integration time and from additions to the dwarf galaxy sample from future surveys. For CTA, our projected sensitivity includes the US contribution and assumes a 500 hour exposure to the Galactic Center. Further details of how the constraints are calculated, including our treatment of e.g. halo profiles, can be found in~\cite{Cahill-Rowley:2014boa}.

\begin{figure}[htbp]
\centerline{\includegraphics[width=3.5in]{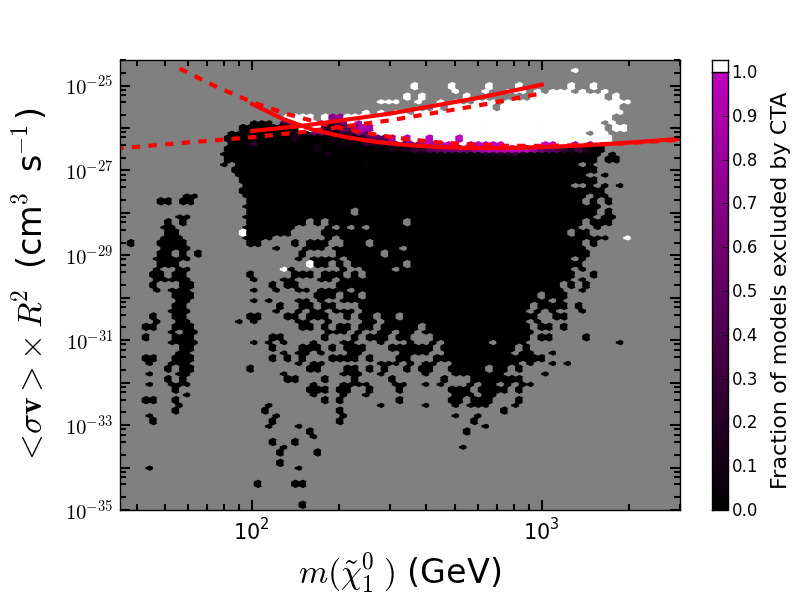}
\hspace{-0.10cm}
\includegraphics[width=3.5in]{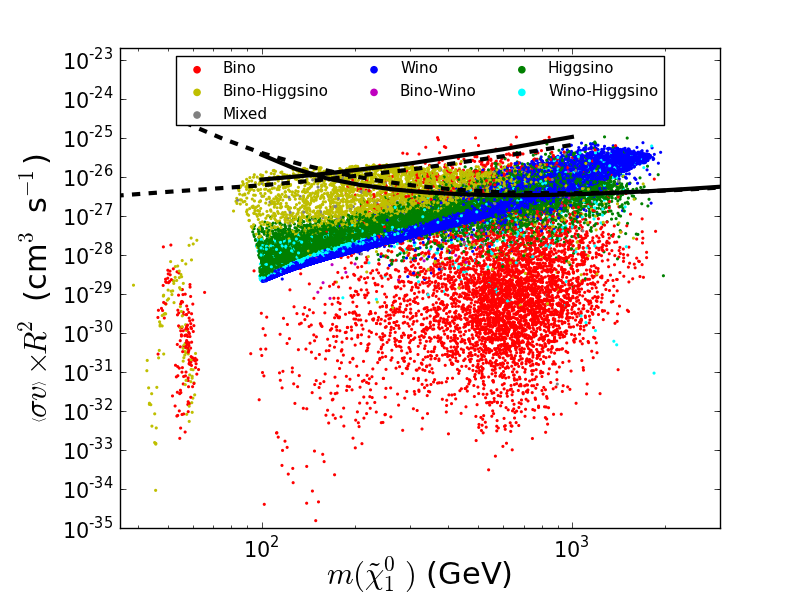}}
\vspace*{-0.10cm}
  \caption{The LSP mass vs scaled annihilation cross section plane, color coded by the fraction of models that could be excluded by CTA (left) and by the LSP composition (right). Red lines
    represent the projected sensitivities for \Fermi~ (peak sensitivity at low masses) and CTA (peak sensitivity at high masses) to DM annihilating exclusively into $b \bar b$ (dashed) and $W^{+} W^{-}$ (solid) final states.}
\label{fig4}
\end{figure}

Figure~\ref{fig4} shows the expected impact of these future measurements on the pMSSM models. In the left panel, we show the LSP mass - annihilation cross section plane, with the cross section now scaled by two powers of the LSP relic density. Comparing the region excluded by CTA with the region that would be excluded if the LSP annihilated only to $b\bar{b}$ or $W^+W^-$ final states, we see that the single-channel limits provide a fairly accurate description of the pMSSM exclusion. Although not shown, the same is true for the \Fermi~ pMSSM exclusion. The right panel of the figure shows the same plane, but now color coded by LSP type. We see that the well-tempered neutralinos predict a strong signature in indirect detection experiments as well as direct detection experiments, but that the $Z$ and $h$ funnel bino-Higgsino mixtures are now well out of reach, as are the heavier co-annihilating and $A$ funnel binos. On the other hand, heavy winos and Higgsinos (which had low scattering rates) are well within the CTA sensitivity due to their large chargino-mediated annihilation rates. Note that light winos and Higgsinos also have large annihilation cross sections, but they suffer from suppressed relic abundances as shown in Figure~\ref{fig00}, making them more challenging to observe.

\begin{figure}[htbp]
\centerline{\includegraphics[width=3.5in]{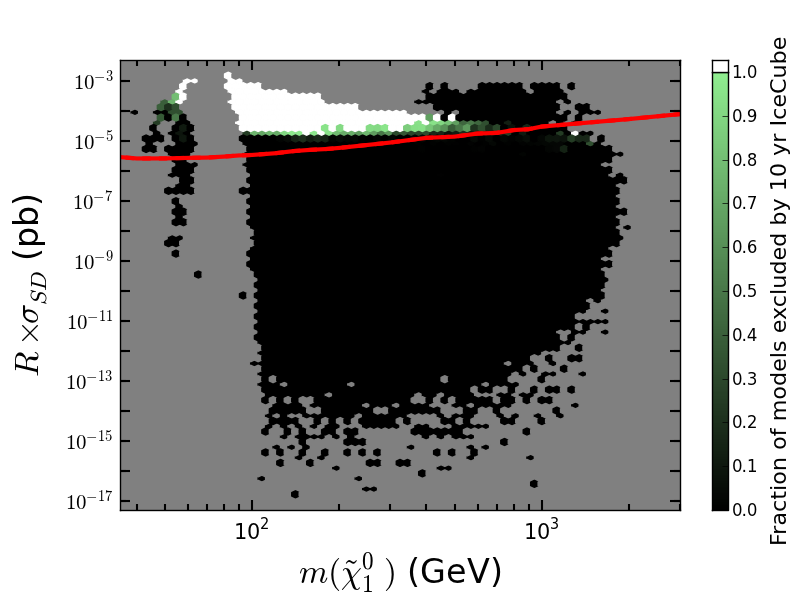}
\hspace{-0.10cm}
\includegraphics[width=3.5in]{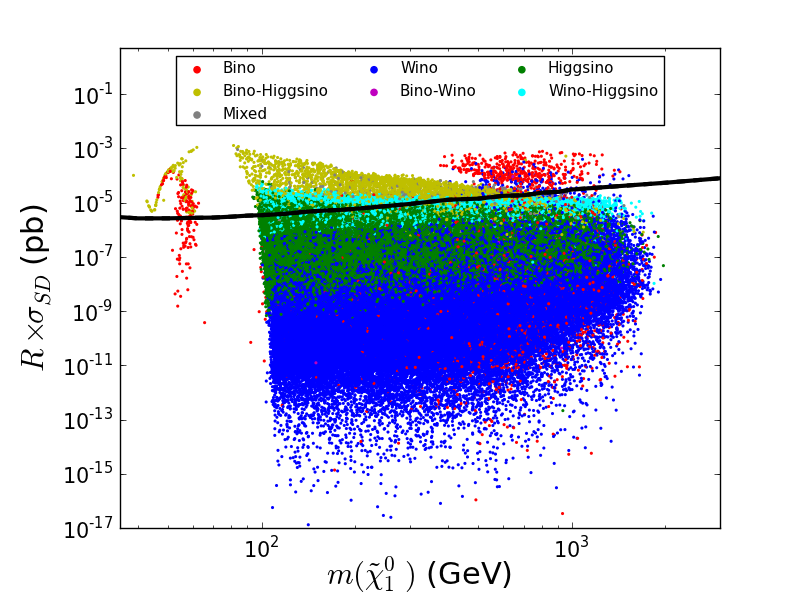}}
\vspace*{-0.10cm}
  \caption{The LSP mass vs SD scattering cross section plane, color coded by the fraction of models that could be excluded by IceCube (left) and by the LSP composition (right). The red line
    represents the projected LZ sensitivity.}
\label{fig5}
\end{figure}

Neutrino telescopes such as IceCube provide another potential discovery channel for WIMP dark matter. In this case, the searches rely on neutralino dark matter being captured by the sun, sinking to the solar core, and annihilating, producing high energy neutrinos either directly or through cascade decays. If the product of capture and annihilation cross sections is large enough, solar capture leads to an equilibrium density of DM particles in the solar core, whose annihilation rate is proportional to the DM-nucleon elastic scattering cross section. The left panel of Figure~\ref{fig5} shows the impact of the projected IceCube sensitivity in the LSP mass vs SD cross section plane. We see that IceCube is only sensitive to models with a large SD cross section (within the reach of LZ), since this is necessary to achieve a large enough capture rate. However, the exclusion is incomplete even for models with large SD cross sections, since the IceCube sensitivity also depends on the annihilation channel, and on the annihilation cross section in models where it is small enough to keep the model out of equilibrium. The right panel shows that the region with the best IceCube sensitivity corresponds mainly to well-tempered neutralinos, as we might expect from the fact that the IceCube signal relies on sizable scattering and annihilation rates.

\section{Complementarity: Putting It All Together} 

\begin{figure}[htbp]
\centerline{\includegraphics[width=3.5in]{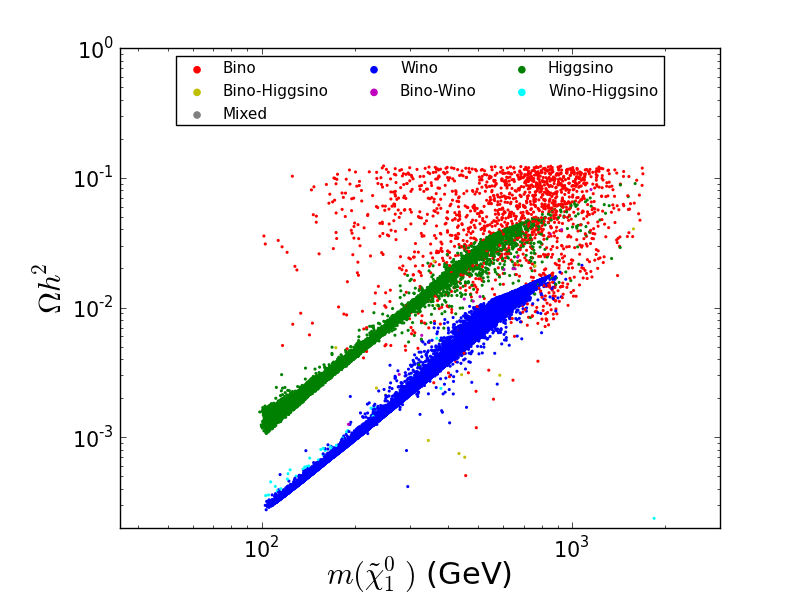}
\hspace{-0.50cm}
\includegraphics[width=3.8in]{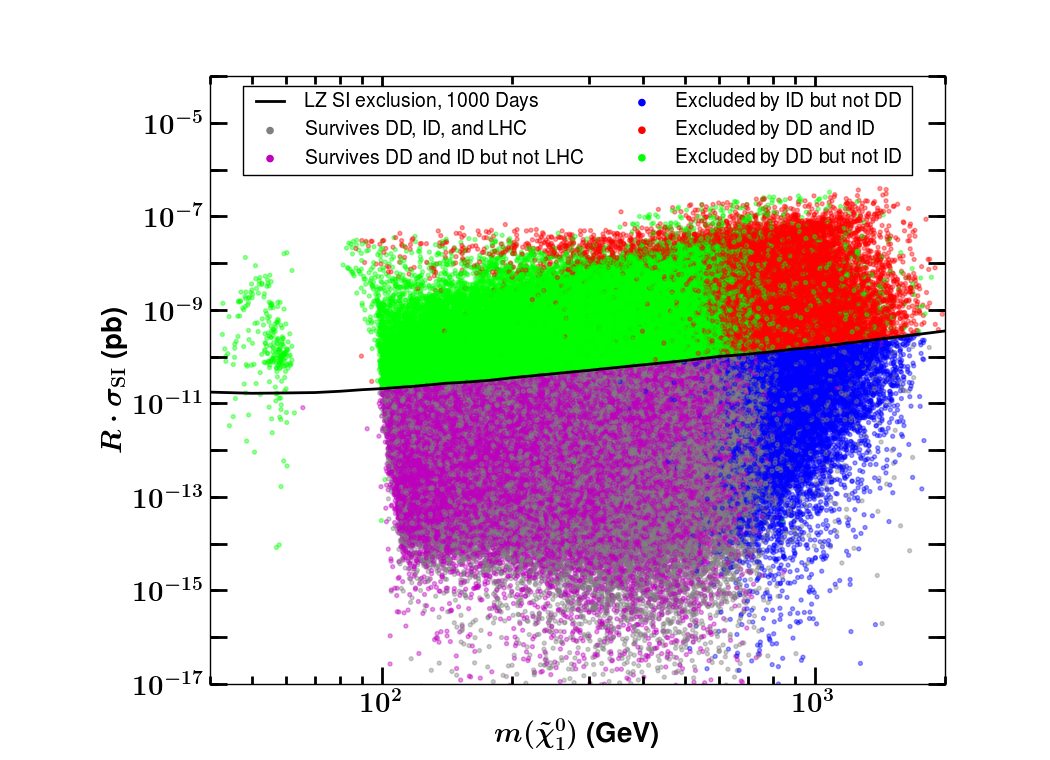}}
\vspace*{-0.10cm}
\caption{Left panel: Thermal relic density as a function of the LSP mass for the pMSSM models surviving after the expected constraints from all the searches are taken into account, color coded by the electroweak properties of the LSP. Compare with Figure~2. Right panel: Comparisons of the sensitivities of the various searches, color coded as indicated, in the LSP mass-scaled SI cross section plane. The anticipated SI limit from LZ is shown as a guide to the eye.}
\label{fig6}
\end{figure}

Now that we have examined the expected sensitivities of flagship experiments in each search category, we can look at the results that we might expect from combining the different experiments. The left panel of Figure~\ref{fig6} shows the analog of Figure~\ref{fig00} after null results from all of the experiments considered here. We see that many regions of the original parameter space have been removed entirely. In particular, bino-like LSPs in the $Z/h$ funnel region have been removed by LZ, while CTA has removed all winos and Higgsinos with a relic abundance approaching the observed DM abundance. All DM experiments considered had a strong sensitivity to the well-tempered neutralinos, which are likewise completely excluded. Interestingly, the only remaining models which saturate the DM abundance are highly-pure binos which coannihilate with other sparticles or annihilate through the A funnel. Both scenarios can potentially be probed by future LHC data, although the sfermion coannihilation region will remain challenging (due to spectrum compression), as will searches for the pseudoscalar $A$ in models with low $\tan \beta$.

The right panel of Figure~\ref{fig6} shows the regions where the different experiments are most sensitive in the LSP mass-SI cross section plane. Here we see the broader patterns of the experimental sensitivities, and their underlying complementarity. In particular, we see that the LHC is sensitive mainly to models with light LSPs, while indirect detection (specifically CTA) is very sensitive to heavy LSPs. Direct detection, by contrast, is essentially independent of the LSP mass. The three experiment classes therefore cover overlapping but orthogonal regions of the parameter space, showing a high degree of complementarity.

\begin{figure}[htbp]
\centerline{\includegraphics[width=3.5in]{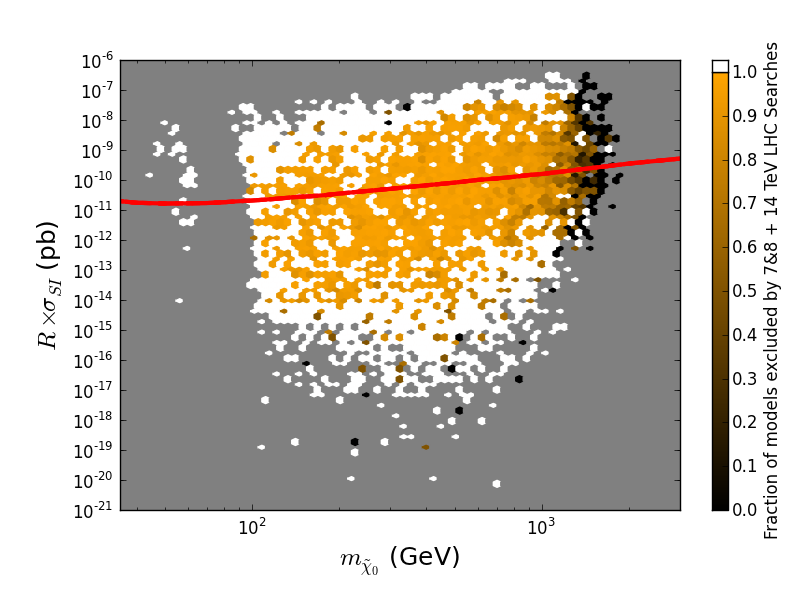}}
\vspace*{-0.10cm}
\caption{The fraction of models with the correct Higgs mass which are excluded by the combination of the 14 TeV jets + MET and the 0$\ell$ + 1$\ell$ stop searches with 300 fb$^{-1}$, shown in 
the LSP mass-scaled SI cross section plane. The red line shows the expected limit on the Xenon SI cross section from LZ.}
\label{14TeVComp}
\end{figure}

Finally, we can ask how the 14 TeV LHC will affect our results. Although we are only able to simulate three of the many searches that will be performed at 14 TeV, our results indicate the qualitative change that may be expected. In particular, we see from Figure~\ref{14TeVComp} that the LHC sensitivity is now expected to have a cutoff for LSP masses of about 1.3 TeV, as opposed to the 700 GeV cutoff seen in the 7/8 TeV results. (Note that Figure~\ref{14TeVComp} shows only models with the correct Higgs mass as a result of the large computational effort required to simulate the high luminosity run). Despite the increased sensitivity to heavy LSPs, however, models without sizable colored production will remain viable regardless of LSP mass. This is demonstrated by the incomplete exclusion at low LSP masses in Figure~\ref{14TeVComp}. Overall, the increase in LHC energy will not change the basic complementarity between the LHC sensitivity, peaking at low LSP masses, and the CTA sensitivity, which extends to very high LSP masses, well beyond the range considered here.

\section{Conclusion} 

After examining the effects of the different search techniques within our pMSSM framework, we conclude that each search category will provide essential sensitivity to important DM scenarios, and that the next generation of experiments will represent substantial progress in our exploration of this space. In particular, critical contributions will be made by the LHC, LZ, and CTA. Although IceCube and \Fermi~ are generally not sensitive to models missed by the other experiments, they would be instrumental in providing an independent confirmation of a signal and in beginning the process of characterizing the DM. We look forward to this new chapter in WIMP searches, and hopefully to the discovery of WIMP DM!

\end{document}